\documentclass[10pt,a4paper,twocolumn]{article}
\usepackage{a4wide,graphicx,cite}
\title{\bf Microfluidic Exploration of the Phase Diagram
of a Surfactant/Water Binary System}

\author{J. Leng{$^{+,\dagger,*}$}, M. Joanicot{$^+$}, A. Ajdari{$^{\dagger}$}}

\date{}
\begin{document}
\maketitle
\noindent {$^{+}$}\emph{\small LOF, unit\'e mixte CNRS/Rhodia/Bordeaux-I, 178 avenue du Dr Schweitzer, F-33608 Pessac Cedex, France.}
{$^\dagger$}\emph{\small Laboratoire Th\'eorie et Microfluidique, UMR 7083 CNRS-ESPCI, 10 rue Vauquelin, F-75005 Paris, France}. {$^{*}$} Corresponding author: \texttt{jacques.leng-exterieur@eu.rhodia.com}.

\begin{abstract}
\sf We investigate the behaviour of a binary surfactant solution (AOT/water) as it is progressively concentrated in microfluidic evaporators. 
We observe in time a succession of phase transitions from a dilute solution up to a dense state, which eventually grows and invades the microchannels. Analyzing these observations, we show that, with a few experiments and a limited amount of material, our microdevices permit a semi-quantitative screening of the equilibrium phase diagram as well as a few kinetic observations.
\end{abstract}

While originally developped in the fields of pharmaceutical and biochemical industries (\emph{via} combinatorial analysis, robotics, etc.~\cite{hobden1992,burbaum1997}), high throughput screening has become a strategy of prime importance for activities requiring the formulation of multicomponents systems (food stuff, cosmetics, etc.)~\cite{fletcher2006}. In this context, the parallelization of tests through the use of miniaturized devices (i.e. microfluidic chips) 
is an attractive avenue to increase productivity.
We have recently introduced microevaporators (devices with integrated pervaporation) for the concentration of small volumes
of aqueous solutions ~\cite{leng2006}, and demonstrated
the great level of control brought up by miniaturization
on very dilute solutions, and in a study of the apparition and growth of crystals in a simple electrolyte (KCl) solution.

In this Letter we go one step further, and show that these devices permit the screening of the properties of more complex systems, here a binary surfactant/water mixture
that exhibits \emph{se\-ve\-ral phase transitions} en route towards high concentrations. We 
indeed report below the successive occurence of the \emph{four} phases expected at equilibrium, and also out-of-equilibrium features related to the densification process. In particular, the concentration process leads eventually
to the nucleation of a dense hexagonal phase which invades the microchannels. We show that this growth is limited by the solute supply, and thus
controlled by the design and operation of the device. A quantitative analysis
yields a determination of the density of the growing hexagonal phase in very good agreement with literature data.

\begin{figure}
\centering
\includegraphics[width = 0.8\linewidth]{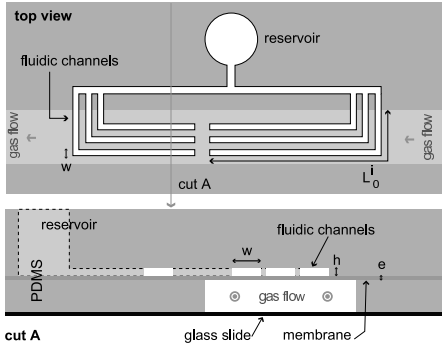}
\caption{\small \sf \label{fig:setup} Schematic top and side views of the microevaporator. 2 sets of dead end channels of different lengths $L_0$ (face to face for design compactness) are connected to a reservoir and positioned above the gas channel (thickness $\approx 0.5$~mm). The fluidic part is molded in PDMS and sealed with a $e\approx 10\,\mu$m thick PDMS membrane. In the real design there are $2\times 24$  channels, with $h=25\;\mu$m, $w=53\;\mu$m, and $L_0$ between 3 and 18~mm).}
\end{figure}

\emph{Device, surfactant system, and experimental protocol - }
The microdevice used here (fig.~\ref{fig:setup}) relies on identical principles
but differs in design from those introduced in reference ~\cite{leng2006},
with a \emph{large evaporation area} (up to $\approx 5\times 5$~cm) allowing a large number of experiments on the same chip. Now standard microfabrication techniques
~\cite{whitesides2001,goulpeau2005} are used to obtain 
two levels of channels in a two-layer poly(dimethylsiloxane) (PDMS) system.
A set of 48 dead-end microchannels (thickness $h\sim 25 \mu m$), originating from a common reservoir, run over the membrane (a thin PDMS layer of thickness $e\sim 10 ~ \mu m$), below which a wide channel allows removal of the water that permeates through this membrane. The channels have a gradually varying length $L_0$ overlying the membrane, which yields a \emph{gradient of concentration rate} ~\cite{leng2006}.

Indeed, if water permeates at a volumetric flow rate $v_e$ through the bottom membrane, the compensating flow from the reservoir has an average velocity $v_0=v_eL_0/h$.
If the reservoir is filled with a dilute solution at concentration $c_0$
the flux of solute in the microchannel is $j_0=c_0v_0$ and thus increases with $L_0$.
To estimate $v_e$ for our device, we connect the reservoir to a transparent tubing of small inner diameter, fill the system with deionized (DI) water, turn evaporation on, and monitor the motion of the meniscus inside the tubing. From the \emph{total} evaporation flux $Q_e$ we deduce the \emph{evaporation velocity} $v_e = Q_e/S_t$, where $S_t$ is the total evaporation surface. We measure $v_e = 52\pm 4$~nm/s. This sets the typical flow velocity $v_0 = \mathcal{O}(10-100\;\mu$m/s),
and an important time scale for the process
$t_e = h/v_e = L_0/v_0 = 460 \pm 50$~s. 

We use aqueous solutions of docusate sodium salt (AOT, $M_w = 444.56$~g/mol, density $\approx 1.1$~g/mL, from Sigma-Aldrih, used as received),
as this system displays many phases~\cite{rogers1969} with boundaries relatively insensitive to temperature in our working conditions $T=25\pm 2^{\circ}$C. A stock solution of AOT in DI water at $(13.3\pm 0.1)$~mM is prepared by dilution and stirring at $T=40^{\circ}$C; the transparent liquid is then left to equilibrate for several days. Further dilutions are prepared by adding DI water.

In a typical experiment, the microsystem is filled with a dilute surfactant solution \emph{at a concentration $c_0$ below its cmc} (a solution of monomeric AOT), and left to evolve under steady evaporation. We monitor optically the evolution of the contents of the channels at \emph{two different length scales}: High magnification optical microscopy with several contrasts to characterize the AOT mesophases, and wide-field observation using a stereo microscope equipped with a large-chip CCD camera to follow the growth kinetics. In both cases, time-lapsed video acquisition at a frequency of 0.1 Hz offers a sufficient temporal resolution on the time scale of the processes we describe below (hours).
\begin{figure}[t]
\centering\includegraphics[width=0.9\linewidth]{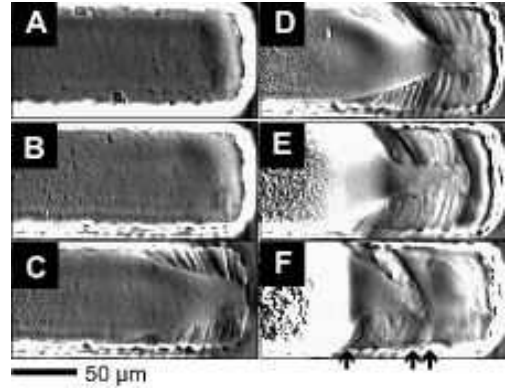}
\caption{\small \sf \label{fig:snapshots} Time-lapsed microscopy of the densification of AOT in water at the tip of a microchannel (phase contrast; initial AOT concentration $c_0 \approx 13$~mM). A. $t = 500$~s: a \emph{fluffy texture} of lamellar blobs and vesicles mostly close to the walls; B. $t=1500$~s: densification of the fluffy texture at the tip yielding a neat structure; C. $t=1750$~s: after nucleation of a mesophase from the sides of the tip; D. $t=2000$~s: growth of this mesophase; E. $t=2250$~s: nucleation of a new mesophase at the tip; F. $t=2750$~s: band strutcture after the nucleation of a last phase (arrows indicate limits between successive textures).}
\end{figure}
\emph{Kinetic generation of phases -}
During the course of the experiment, several stages take place that are best viewed at high magnification (see also supplementary material). Starting from a perfectly transparent liquid, we first observe the occurence of small and mobile objects ($<5 \mu$m, fig.~\ref{fig:snapshots}A, likely to be vesicles as they are not visible without phase contrast). Then, blobs and giant vesicle like structures grow \emph{from the walls} (fig.~\ref{fig:snapshots}A). They are also present in solution, some of them being teared away from the walls and advected towards the tip of the channel. This type of rough dispersion is typical of low solubility surfactants (lipids, AOT, ...) and is prone to kinetic effects~\cite{buchanan1998,leng2003}. The dispersion then progressively densifies, loosing at some stage its rough aspect (fig.~\ref{fig:snapshots}B). Soonafter, a nucleation event occurs (fig.~\ref{fig:snapshots}C): two slabs symmetrically grow from the corners of the channel's tip; they do not merge but join through \emph{domains} typical of a \emph{lamellar phase} (oily streaks and maltese crosses~\cite{chandra1986}, fig.~\ref{fig:snapshots}D). After another latency time, during which the lamellar slabs grow, a new phase shows up at the tip of the channel (fig.~\ref{fig:snapshots}E). Eventually, \emph{a final texture appears}, and grows "pushing" away a front bearing 
bands of the previously observed phases (fig.~\ref{fig:snapshots}F).

\begin{figure}[t]
\centering 
\includegraphics[width=0.8\linewidth]{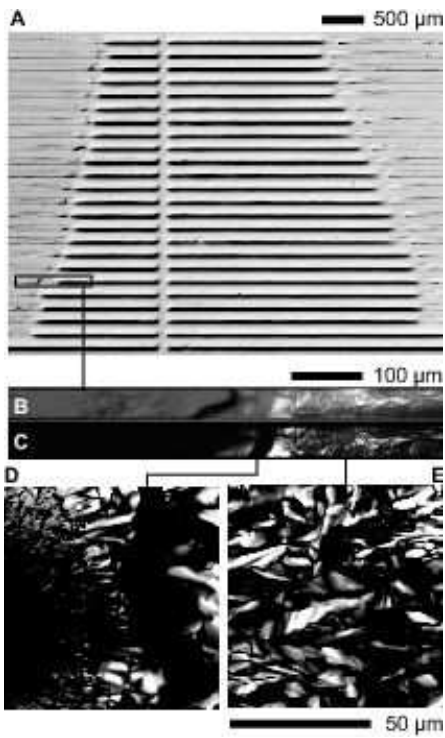}
\caption{\small \sf \label{fig:all_scales} Overview of the growth kinetics. A: Large scale view of the parallelized device during the growth of the densest phase (the very contrasted zone) after $\approx 4$~hours for an initial AOT concentration
$c_0 = 13.3$~mM. The growth rate increases with channel length (the shortest is in the upper left, the longest is the bottom right). B, C: Magnification of the front zone (B. partially crossed polarizers, C. totally crossed polarizers) between a fluffy texture and the dense phase, showing \emph{a band structure} through birefringence properties of successive phases. D, E: higher magnification under crossed polarizers.}
\end{figure}

At a larger scale, only parts of the kinetics can be tracked: we monitor the occurence of a fluffy texture over a finite channel length, which later collapses and the subsequent nucleation and growth of a dense phase, which has a deep and specific contrast in bright field view (fig.~\ref{fig:all_scales}A). This picture clearly illustrates the link between channel length and growth rate: the longer, the quicker. 
Pictures B and C of fig.~\ref{fig:all_scales} show a magnification of the "band structure" of the growing front, observed under (partially and totally) crossed polarizers, with an alternation of \emph{isotropic and birefringent bands}. With finer microscopy (fig.~\ref{fig:all_scales}D and E), we observe typical textures of a lamellar phase (D) and a hexagonal phase (E). 

\begin{figure}[t]
\centering 
\includegraphics[width=0.8\linewidth]{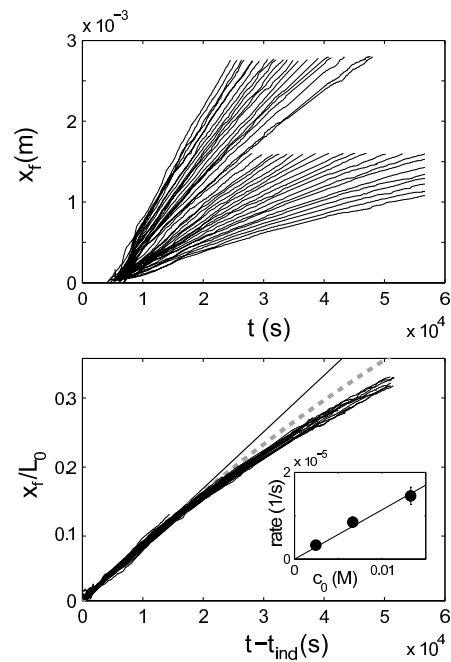} 
\caption{\small \sf \label{fig:growth} Top: front of the growing dense phase $x_f$ against time, in 48 microchannels of different lengths $L_0$. Bottom: same results for $x_f/L_0$ against reduced time ($t_{ind}$ is the growth start time); Inset: initial slope $\dot{x}_f/L_0$ against reservoir concentration $c_0$. The full and dashed lines are linear and exponential fits (see text).}
\end{figure}

This sequence of phases is perfectly compatible with the AOT phase diagram, which 
consists of (increasing surfactant contents) a monomeric state, a lamellar, a cubic, and a hexagonal phase, with a large miscibility gap between the solution and the lamellar phase that reflects the very low solubility of AOT. We identify here three states (lamellar suspension, lamellar phase, hexagonal phase) out of four; the last one is a concentrated phase lying between the lamellar and the hexagonal phases, and is isotropic. The cubic phase of the phase diagram is a perfect candidate, but only \emph{in situ} X-ray scattering could definitively assess it. 

We thus infer that the surfactant accumulation process yields a \emph{qualitative view on the phase diagram},  through the successive nucleation events  and \emph{the band structure} during the ultimate growth. Recall however that our experiments provide a kinetic view, a feature of prime importance for a correct mapping of the concentration kinetics onto the phase diagram.

\emph{Growth kinetics -} We now show how
analysis of the kinetics permits the quantitative determination
of a phase boundary for this complex systems, applying a recipe succesful for electrolyte solutions ~\cite{leng2006}.

We monitor systematically the eventual growth of the hexagonal phase with image processing. A single experiment provide 48 growth kinetics (fig.~\ref{fig:growth}, top), noticeably dependent on the length of each channel $L_0$. The growth rate increases with $L_0$ and decreases slightly with time. A remarkable collapse
is obtained if the front position is normalized by the channel length $L_0$ (fig.~\ref{fig:growth}, bottom). Performing experiments at different concentrations, we find
the initial reduced growth rate $\dot{x}_f/L_0$ proportionnal to $c_0$ (inset of fig.~\ref{fig:growth}, bottom). This assesses our operational
control of the kinetics through both $L_0$ and $c_0$.

This dependence is consistent with mass conservation, assuming that the growth is limited by the solute supply at $j_0=c_0v_0=c_0L_0/t_e$
from the reservoir. If the dense phase grows at a fixed concentration
 $c_d$, a \emph{constant} growth rate $\delta x_f/\delta t = (c_0/c_d) L_0/t_e$  is expected. Fitting this law to the early growth stage (thin solid line, fig.~\ref{fig:growth}, bottom), and using our separate measurement of $t_e$, we get an estimate for the concentration of the newly born dense state: $c_d = 2.2 \pm 0.2$~M, or  $84\pm 8\%$in mass of AOT,
remarkably close to the lower bound of the hexagonal phase of AOT in water ($\approx 83\%$~\cite{rogers1969}).

We also observe a progressive slowing down in the growth,
and that eventually the dense state \emph{does not grow beyond the evaporation zone}, which suggests that \emph{evaporation from the dense phase is weak}. Assuming that is strictly zero leads an effective evaporation length $L(t) = L_0 -x_f(t)$ and a solute flux that decrease as the dense state grows, and predicts an exponential 
slowing down: $x_f(t)/L_0 = ( 1 - e^{-\frac{c_0}{c_d} \frac{t}{t_e}})$. This accounts
for most but not all of the growth rate decrease (thick dashed line of fig.~\ref{fig:growth}, bottom).
Several factors may alter this simple exponential prediction. In particular, as the dense hexagonal phase exists over a range of concentrations ($\approx 80  \to 100 \%$ AOT w/w), the corresponding band may grow more slowly because it becomes denser
in time, possibly up to a pure AOT solid.
Such a final densification is possible if a weak evaporation persists within the densest phase, inducing in this phase a permeation flow which generates osmotic compression. 

\emph{Conclusion -}
For the \emph{continuous screening} goal pursued here, microevaporators are powerful and accurate tools that offer a semi-quantitative description of a surfactant \emph{phase diagram} with a limited set of experiments, of material, and of analysis. The observed kinetic pathway of AOT concentration in water provides a general scenario for the densification mechanism: a journey towards high concentration, crossing on the way every state, and preserving this memory in the \emph{band structure} of the concentration gradient during growth.
This experimental control in the \emph{manipulation of solute at small scale}---especially for its kinetic aspect--- sets firm grounds for further studies of more complex mixtures, and of \emph{kinetic aspects} of the concentration process and the intricate physics involved. 

\emph{Acknowledgements -}
We thank \emph{R{\'e}gion Aquitaine} for funding and support.

\emph{Supporting Information Available -} Two videos showing the concentration process are supplied as supplementary material.
This material is available free of charge via the Internet at \texttt{http://pubs.acs.org}.

\small
\bibliography{evap2}
\bibliographystyle{unsrt}
\end{document}